# PROTON RADIUS PUZZLE AND LARGE EXTRA DIMENSIONS

LI-BANG WANG[*]

*Department of Physics, National Tsing Hua University, Hsinchu,30013,Taiwan*
*lbwang@phys.nthu.edu.tw*

WEI-TOU NI

*Department of Physics, National Tsing Hua University, Hsinchu,30013,Taiwan*
*weitou@gmail.com*

We propose a theoretical scenario to solve the proton radius puzzle which recently arises from the muonic hydrogen experiment. In this framework, $4 + n$ dimensional theory is incorporated with modified gravity. The extra gravitational interaction between the proton and muon at very short range provides an energy shift which accounts for the discrepancy between spectroscopic results from muonic and electronic hydrogen experiments. Assuming the modified gravity is a small perturbation to the existing electromagnetic interaction, we find the puzzle can be solved with stringent constraint on the range of the new force. Our result not only provides a possible solution to the proton radius puzzle but also suggests a direction to test new physics at very small length scale.

## 1. Introduction

The charge radius of the proton has drawn a lot of theoretical interest since a very precise Lamb shift measurement has been performed on muonic hydrogen.[1] The measured rms charge radius $r_p = 0.8418(7)$ fm is significantly smaller than the CODATA value of $0.8775(39)$ fm, which is a combined result from electron scattering experiments and atomic hydrogen spectroscopy.[2] A recent measurement on the hyperfine transition in muonic hydrogen yields a result of $r_p = 0.84087(39)$ fm and reinforces the proton radius puzzle.[3] Numerous investigations have been attempted to explain this puzzle and they include, for example, multi-photon exchange correction to the electron scattering results[4], peculiar electromagnetic form factors that may mislead the interpretation of the scattering data[5] and differences in interaction between muon and electron[6-7] which directly imply physics beyond standard model. Pohl and coworkers have recently examined all experimental methods together with theoretical models[8] and conclude the proton radius puzzle is real. Although the existence of new physics in muonic atoms seems unlikely, it cannot be ruled out. Previously, we have looked into the empirical constraints on extra dimension theories using atomic transition frequencies of hydrogen and muonium.[9] Similar ideas of using spectroscopy to constrain/probe extra dimensions/gravitational

---

physics have also been studied in various places.[10-12] In this paper, we look into possible gravitational interactions based on large extra dimensions (LEDs)[13-19] for a possible solution of the proton size puzzle. Specifically we adapt the theory by Arkani-Hamed, Dimopoulos and Dvali[16] (known as the ADD model) to obtain the energy shift in atomic transitions. ADD model is proposed to explain why gravity is so weak compared with the other three known interactions. We found that it is possible to give an energy contribution of 0.31 meV (75 GHz) to muonic hydrogen 2P → 2S transition from some ADD models to explain the proton radius puzzle. In section 2, we review atomic energy levels and nuclear radii. In section 3, we adapt the ADD models to calculate the relevant energy shift. In section 4, we summarize with discussions.

## 2. Atomic Energy Levels and Nuclear Radii

The non-zero nuclear size could affect the atomic transition frequencies due to distortions of the Coulomb potential when the electron has the probability inside the nucleus. This volume effect has been known as field shift in atomic transitions and formed the basis for spectroscopic determination of the nuclear charge radii. For atomic hydrogen, the field shift is (in SI unit):

$$\Delta E = \frac{e^2}{6\varepsilon_0} |\psi(0)|^2 \cdot r_p^2 \tag{1}$$

Where $\psi(0)$ is the wavefunction of the electronic state of interest at origin. Providing that all other contributions including QED and hadronic effects for atomic energy levels can be calculated very precisely, the difference between measured transition frequencies and theoretical calculations can then be attributed to finite nuclear size. This method has been widely used in simple atomic systems where the wavefunction can be calculated precisely and valuable results are obtained using high-resolution laser spectroscopy.[20-22] In the case of muonic atom, the effect of field shift is much larger since muon is 207 times heavier than electron and is several million times more likely to appear inside the proton. However, since the muon's Bohr radius is only about 300 times larger than the proton radius, it is not quite accurate to assume the muon's wavefunction is constant across the proton's volume. As a result, higher order effect and the proton's charge distribution have to be considered. The predicted $^2S_{1/2, F=1}$ to $^2P_{3/2, F=2}$ energy difference is given in Ref. 1

$$\Delta E = 209.9779(49) - 5.2262 \cdot r_p^2 + 0.0347 \cdot r_p^3 \text{ meV} \tag{2}$$

where $r_p$ is given in fm. The discrepancy arises when the measured result is different from the predicted value using Eq. (2) and with $r_p$ = 0.8775(39) fm from CODATA. Namely, 0.31 meV or 75 GHz higher in transition frequency is obtained. No other effects can be found so far to produce such large energy shift. Since the results from electron scattering experiments and hydrogen atomic spectroscopy are in reasonable agreement, the discrepancy has bothered physicists for years after the result was announced and is the so-called proton radius puzzle.

To solve this puzzle, we introduce an extra interaction $V_{LED}$ due to modified gravity within very short range between the proton and lepton. Since the interaction is attractive, it provides extra binding for the 2S lepton while the 2P lepton receives negligible effect due to vanishing wave function at origin. If we assume $V_{LED}$ acts as a small perturbation, the energy shift due to the new interaction can be estimated using perturbation theory and is set to be 0.31 meV to make up the higher measured transition energy.

$$\Delta E_{LED}(2P \rightarrow 2S) = \int \psi_{2P}^* V_{LED} \psi_{2P} d\tau - \int \psi_{2S}^* V_{LED} \psi_{2S} d\tau = 0.31 \text{ meV} \quad (3)$$

where $\psi_{2S} = (8\pi)^{-1/2}(a_0)^{-\frac{3}{2}} \cdot e^{-\frac{r}{2a_0}}(1 - r/2a_0)$, $a_0 = 2.85 \times 10^{-13}$ m is the Bohr radius of the muon. By choosing appropriate integration limits, the parameters for $V_{LED}$ can then be determined. In the following section, we use ADD model to construct the extra interaction.

## 3. Large Extra Dimensions and ADD Model

Gravity theory with large extra dimensions provides such energy shift in the right direction. For muonic hydrogen, the gravitational binding of muon to the proton is larger than that for electron in the hydrogen. The ADD model used here is first proposed by Arkani-Hamed, Dimopoulos and Dvali[15] to solve the gauge hierarchy problem for gravity. Suppose there are $n$ extra dimensions of radius $\sim R_n$ and only gravity can propagate into extra dimensions, the gravity at very short range can be much stronger than the normal gravity. Two test masses of mass $m_1$, $m_2$ placed within a distance $r < R_n$ will feel an attractive gravitational potential which can be constructed using $(4 + n)$ dimensional Gauss's law and is written explicitly as

$$V_{LED}(r) = -\frac{m_1 m_2}{M_{pl(4+n)}^{n+2}} \frac{1}{r^{n+1}}, \quad (r < R_n)$$
$$V_{LED}(r) = -\frac{m_1 m_2}{M_{pl(4+n)}^{n+2} R_n^n} \frac{1}{r}, \quad (r > R_n) \quad (4)$$

Continuity condition gives

$$M_{pl}^2 = M_{pl(4+n)}^{n+2} \cdot R_n^n \quad (5)$$

where $M_{pl} = 1.22 \times 10^{19}$ GeV/c$^2$ is the Planck mass scale. From Eq. (5), we have the following relation between $R_n$ and the mass scale $M_{pl(4+n)}$

$$R_n \sim \pi^{-1} \, 10^{-19+(32/n)} (1\text{TeV}/M_{pl(4+n)})^{1+(2/n)} \text{ m}. \quad (6)$$

In the ADD models with $n$ extra dimensions, the mass scale $M_{pl(4+n)}$ and the extra dimension radius are related. Their exact relation depends on the topology and structure of the extra dimensions, i.e. the details of the theories.

We can now calculate the extra binding energy for muonic hydrogen with this modified gravitational potential using Eq. (3). The integration limit is set according to the following argument. The upper limit is $R_n$ as one can expect since the modified gravity only exists within this radius. Newtonian gravity above this range contribute negligible energy shift and is usually neglected. The lower limit corresponds to a cutoff range within which the muon cannot penetrate and it represents a hard core effect.[23] In our simple phenomenological approach, there are two parameters – the mass scale (equivalently, the size $R_n$ of compact dimensions) and the hard core cutoff radius. The physics will depend on how the topology of compact dimensions and how they are curved (e.g., flat, with constant curvature or else). To simplify things, we just use a cutoff radius and this treatment is also often utilized in other fields of physics. For example, in calculation of nuclear structure, hard core of nucleons with parameterized radius is used to represent the repulsive forces at short distance and prevent the nuclear matter from being too dense.[24] In N-body simulations of astronomical objects, a maximum force between two objects is assumed so that the interaction will not become divergent at zero range.[25] In theories with extra dimensions, brane-crystals have been utilized to generate large extra dimensions and the $D$-branes experience a van der Waals force combined with a hard core repulsive interaction and lead to stable lattices.[26-28] In the evaluation of $R_n$, we use simple hard core models with different radii. The values include the proton radius scale $\sim 10^{-15}$ m, the electroweak length scale $10^{-18} \sim 10^{-19}$ m and the smallest length scale that LHC can probe. The proton radius is a natural choice of the cutoff range if we assume the proton is a uniform sphere and thus the gravitational interaction quickly decreases if the lepton is inside the proton. The electroweak length scale is also used according to the ADD model that the gravity should switch off theoretically at distances below this scale. The last one is the limit of the quark size. This represents the possible quark structure and can be probed at LHC in the future. The results are shown in Table 1.

Table 1. The size $R_n$ and mass scale $M_{pl(4+n)}$ of LED which gives correct energy contribution of 0.31 meV to solve the proton radius puzzle in hard core models with core radius $r_0$ of 1 fm, 1 am, 0.1 am and 0.02 am. The selection of the various hard core radii is explained in the text.

| $n$ | | Size ($R_n$) in m and mass scale $M_{pl(4+n)}$ in TeV/c² of the LED | | | |
|---|---|---|---|---|---|
| | | $r_0 = 1$ fm | $r_0 = 1$ am | $r_0 = 0.1$ am | $r_0 = 0.02$ am |
| 3 | $R_n$ | $5.0\times10^{-4}$ | $4.8\times10^{-5}$ | $2.2\times10^{-5}$ | $1.3\times10^{-5}$ |
| | $M_{pl(4+n)}$ | $4.8\times10^{-4}$ | $2.0\times10^{-3}$ | $3.1\times10^{-3}$ | $4.3\times10^{-3}$ |
| 4 | $R_n$ | $6.8\times10^{-7}$ | $2.2\times10^{-8}$ | $6.8\times10^{-9}$ | $3.1\times10^{-9}$ |
| | $M_{pl(4+n)}$ | $2.8\times10^{-4}$ | $2.8\times10^{-3}$ | $6.0\times10^{-3}$ | $1.0\times10^{-2}$ |
| 5 | $R_n$ | $1.3\times10^{-8}$ | $2.0\times10^{-10}$ | $5.1\times10^{-11}$ | $1.9\times10^{-11}$ |
| | $M_{pl(4+n)}$ | $1.9\times10^{-4}$ | $3.7\times10^{-3}$ | $9.9\times10^{-3}$ | $2.0\times10^{-2}$ |
| 6 | $R_n$ | $9.0\times10^{-10}$ | $8.7\times10^{-12}$ | $1.9\times10^{-12}$ | $6.4\times10^{-13}$ |
| | $M_{pl(4+n)}$ | $1.5\times10^{-4}$ | $4.7\times10^{-3}$ | $1.5\times10^{-2}$ | $3.3\times10^{-2}$ |
| 7 | $R_n$ | $1.3\times10^{-10}$ | $9.2\times10^{-13}$ | $1.8\times10^{-13}$ | $5.6\times10^{-14}$ |
| | $M_{pl(4+n)}$ | $1.2\times10^{-4}$ | $5.7\times10^{-3}$ | $2.0\times10^{-2}$ | $5.9\times10^{-2}$ |

## 4. Summary and Discussion

In summary, proton radius puzzle may be explained by the energy contribution of 75 GHz to muonic hydrogen 2P → 2S transition from exotic interaction. Modified gravity theories with large extra dimensions may give such exotic contributions. In Table 1, the cases for the hard core radius $r_0 = 1$ fm ($n$ = 3-7) are ruled out by the hydrogen and muonium spectroscopic data.[9] All other $n = 3$ cases are also inconsistent with the muonium bounds.

In Table 1, the mass scale $M_{pl(4+n)}$ are calculated from Eq. (6) for ADD models. The present lower limit from ATLAS on the mass scale of large extra dimensions is 2 TeV.[29] As we can see, the ADD model is no longer tenable as it is. However, here we treat it as a phenomenological framework for exploration. We have also performed the calculation for muonic deuterium and found the energy shift of the same transition is 0.73 meV (176 GHz), which is mainly due to the stronger gravitational force from the deuteron and also from slightly reduced Bohr radius in the muonic deuterium system. More spectroscopic experiments on muonic hydrogen, muonic deuterium and muonium are important to test this scenario. We also emphasize here that even the proton radius puzzle is resolved in the future for other reasons, our analysis can still serve as putting stringent constraints on the dimension scales of modified gravity due to large extra dimensions.

**Acknowledgments**

We are grateful to the National Science Council for financial support (Grant No. NSC101-2112-M-007-007). This work was started for an invited talk "Atomic transition frequencies and extra dimensions" in the Workshop on Interactions between Experimental and Theoretical Efforts for Probing the Microscopic Origin of Gravity and Related Topics, Hsinchu, January 26-27, 2013 organized by the National Center for Theoretical Sciences (NCTS).

**References**

1. R. Pohl *et al.*, *Nature* **466,** 213 (2010).
2. P. J. Mohr, B. N. Taylor and D. B. Newell, *Rev. Mod. Phys.* **84,** 1527 (2012).
3. A. Antognini *et al.*, *Science* **339,** 417(2013).
4. J. Arrington, arXiv:1210.2677.
5. A. De Rújula, *Phys. Lett. B* **693,** 555 (2010).
6. V. Barger *et al.*, *Phys. Rev. Lett.* **106,** 153001 (2011).
7. V. Barger *et al.*, *Phys. Rev. Lett.* **108,** 081802 (2012).
8. R. Pohl, R. Gilman, G. A. Miller and K. Pachucki, arXiv:1301.0905.
9. Z.-G. Li, W.-T. Ni and A. P. Patón, *Chin. Phys.* B **17,** 70 (2008).
10. F. Luo and H. Liu, *Int. J. Theor. Phys.* **46,** 606 (2006).
11. R. Onofrio, *Eur. Phys. J. C* **72,** 2006 (2012).
12. R. Onofrio, *Mod. Phys. Lett. A* **28,** 1350022 (2013).
13. I. Antoniadis, *Phys. Lett. B* **246,** 377 (1990).
14. E. Witten, *Nucl. Phys. B* **471,** 135 (1996).
15. J. D. Lykken, *Phys. Rev. D* **54,** 3693 (1996).

16. N. Arkani-Hamed, S. Dimopoulos and G. Dvali, *Phys. Lett. B* **429,** 263 (1998).
17. K. R. Dienes, E. Dudas and T. Gherghetta, *Nucl. Phys. B* **557,** 25 (1999).
18. S. Chang, S. Tazawa and M. Yamaguchi, *Phys. Rev. D* **61,** 084005 (2000).
19. K. R. Dienes, E. Dudas and T. Gherghetta, *Phys. Rev. D* **62,** 105023 (2000).
20. C. Schwob *et al.*, *Phys. Rev. Lett.* **82,** 4960 (1999).
21. L.-B. Wang *et al.*, *Phys. Rev. Lett.* **93,** 142501 (2004).
22. R. Sánchez *et al.*, *Phys. Rev. Lett.* **96,** 033002 (2006).
23. F. Wilczek, *Nature* **445**, 156 (2007).
24. S. Pieper and R. B. Wiringa, *Annu. Rev. Nucl. Part. Sci.* **51**, 53 (2001).
25. D. Merritt, *AJ* **111**, 2462 (1996).
26. I. Z. Rothstein, *Phys. Rev. Lett.* **110,** 011601 (2013).
27. N. Arkani-Hamed, S. Dimopoulos and J. March-Russell, *Phys. Rev. D* **63**, 064020 (2001).
28. S. Corley and D. A. Lowe, *Phys. Lett. B* **505**, 197 (2001).
29. G. Aad *et al.*, *Phys. Rev. Lett.* **110,** 011802 (2013).